\documentclass[10pt,twocolumn,letterpaper]{article}

\usepackage[pagenumbers]{cvpr} 

\definecolor{cvprblue}{rgb}{0.21,0.49,0.74}
\usepackage[pagebackref,breaklinks,colorlinks,allcolors=cvprblue]{hyperref}

\usepackage[table]{xcolor}
\usepackage{booktabs}
\usepackage{nicematrix}
\usepackage{pifont} 
\usepackage{float}
\definecolor{defaultColor}{RGB}{240,240,240}

\def\paperID{} 
\def\confName{CVPR}
\def\confYear{2026}

\newcommand{\ModuleName}{\textbf{\textit{FoleyDirector}}}
\newcommand{\FeatureName}{Structured Temporal Scripts}
\newcommand{\Feature}{STS}
\newcommand{\AdaName}{Script-Guided Temporal Fusion Module}
\newcommand{\Ada}{SG-TFM}
\newcommand{\BiName}{Bi-Frame Sound Synthesis Framework}
\newcommand{\Bi}{Bi-Frame}
\title{FoleyDirector: Fine-Grained Temporal Steering for Video-to-Audio Generation via Structured Scripts}

\author{
    You Li$^{1,2,3}$ \qquad
    Dewei Zhou$^{1,2}$ \quad
    Fan Ma$^{1,2}$ \quad
    Fu Li$^3$ \quad
    Dongliang He$^3$ \quad
    Yi Yang$^{1,2,\dag}$ \\
    {\small $^1$ The State Key Lab of Brain-Machine Intelligence, Zhejiang University $^2$ ReLER, CCAl, Zhejiang University, China } \\
    {\small $^3$ Intelligent Creation, ByteDance, China  $^\dag$ Corresponding author} \\
    {\tt\small \{uli2000, zdw1999, mafan, yangyics\}@zju.edu.cn}, {\tt\small lifu.01@bytedance.com}, {\tt\small hedlcc@126.com} \\
}

\begin{document}
\maketitle
\begin{abstract}
Recent Video-to-Audio (V2A) methods have achieved remarkable progress, enabling the synthesis of realistic, high-quality audio. However, they struggle with fine-grained temporal control in multi-event scenarios or when visual cues are insufficient, such as small regions, off-screen sounds, or occluded/partially visible objects.
In this paper, we propose \ModuleName, a framework that, for the first time, enables precise temporal guidance in DiT-based V2A generation while preserving the base model’s audio quality and allowing seamless switching between V2A generation and temporally controlled synthesis. \ModuleName~introduces Structured Temporal Scripts(STS), a set of captions corresponding to short temporal segments, to provide richer temporal information. These features are integrated via the Script-Guided Temporal Fusion Module, which employs Temporal Script Attention to fuse STS features coherently. To handle complex multi-event scenarios, we further propose Bi-Frame Sound Synthesis, enabling parallel in-frame and out-of-frame audio generation and improving controllability.
To support training and evaluation, we construct the DirectorSound dataset and introduce VGGSound-Director and DirectorBench. Experiments demonstrate that \ModuleName~substantially enhances temporal controllability while maintaining high audio fidelity, empowering users to act as Foley directors and advancing V2A toward more expressive and controllable.
\end{abstract}    

\section{Introduction}
\label{sec:intro}

Recent advances in AIGC technologies have make greate progress~\cite{flux, migc, migc++, dreamfoley, seedream2025seedream, stablediffusion, wu2025omnigen2, difffoley, hunyuanfoley, mmaudio, ipcir, li2024anysynth, vistallama,zhou20243dis,zhou20253disflux,zhou2025dreamrenderer,zhou2025bidedpo,xu2025contextgen}, enable the generation of high-quality video sequences with remarkable visual fidelity~\cite{wan, cogvideox, hunyuanvideo}. However, most open-source video generation frameworks still produce silent videos, limiting the realism and immersion of generated content.
This absence of sound requires experienced foley artists painstakingly design and record sounds to match every visual detail, making the process labor-intensive and costly.
To address this, the Video-to-Audio (V2A) task aims to restore sound to silent videos by automatically generating high-quality, semantically and temporally consistent audio from multimodal cues. Building upon recent advances in image generation~\cite{flux}, modern V2A models~\cite{mmaudio, hunyuanfoley, kling} have incorporated the DiT architecture, leveraging its joint modeling capability to fuse video–text features, enhance temporal alignment via Synchformer~\cite{synchformer} features, and generate audio with diffusion or flow-matching frameworks.

Although current V2A models can synthesize relatively high-quality audio, they still face several limitations:
\textbf{1) Limited Understanding of multi-event captions.} Existing models treat captions as coarse semantic cues. When describing multiple in-frame and out-of-frame sound events, they struggle to capture the full semantics of each event and their temporal relationships.
\textbf{2) Limited Performance with insufficient visual cues.} As shown in Fig.~\ref{fig:show}, when generating sounds correspond to small pixel regions, occur our-of-frame, are occluded or partially visible, current V2A models have difficulty controlling when these sounds occur~\cite{visual}.
\textbf{3) Limited Controllability.} As V2A generators improve in quality, users increasingly expect to act as \textbf{\textit{Foley Directors}}, yet existing methods provide no effective mechanisms to control or guide the generation process.
These limitations lead us to consider: \textit{Can we provide additional temporal information, supplement insufficient visual cues and enable finer temporal control over audio generation?}

We thus propose \ModuleName, a framework that extends the temporal controllability of DiT-based V2A generation. It consists the following components: \textbf{(i)} We propose Structured Temporal Scripts (STS), a set of captions each corresponding to a short temporal segment of the audio, and design pipeline to extract STS, construct the DirectorSound dataset. \textbf{(ii)} We propose the Script-Guided Temporal Fusion Module (SG-TFM), an adapter that integrates STS features via Temporal Script Attention and employs interleaved RoPE to fuse them into the model in a temporally coherent manner.
\textbf{(iii)} We propose \BiName to handle complex in-frame/out-of-frame scenarios, leverage multimodal control to jointly render both types of sounds and enhance controllability. 
\ModuleName~preserves the quality of the pretrained generator while offering flexible temporal control, and it allows users to seamlessly revert to standard V2A generation by simply disabling the SG-TFM.

We evaluate our model on two benchmarks.
First, we manually constructed DirectorBench includes cases of temporal control and counterfactual control, assess the controllability of our method. Compared with baseline methods, our approach demonstrates significantly stronger controllability, surpassing existing V2A models (raise F1-score from 0.2451 to 0.4819). 
Furthermore, we evaluate audio generation quality on the VGGSound-Director benchmark, where we annotate a subset of VGGSound and synthesize audio for comparison with other models. Our method preserves the distribution of generated audio while achieving lower KL-divergence and higher ISC scores, demonstrating superior fidelity and quality.

\noindent Our contributions are summarized as follows:

\begin{itemize}
  \item [1)] We propose \ModuleName, a framework, for the first time, introduces temporal control in the V2A task, and keeps basic V2A quality.
  
  \item [2)] We decompose complex captions into \FeatureName~and integrates them into model with \AdaName~in a temporally coherent manner, with the \BiName~enables flexible and precise audio control.

  \item [3)] We propose a data annotation pipeline and construct the \textit{DirectorSound} dataset. We also create \textit{VGGSound-Director} and \textit{DirectorBench} to evaluate our method.
  
  \item [4)] We conduct a series of experiments demonstrates that our method significantly improves controllability while preserving the fundamental generation quality.
\end{itemize}
\section{Related Work}
\label{sec:related}

\subsection{Text-to-Audio}
Recent progress in text-to-audio generation has been driven by both Transformer-based and diffusion-based approaches. Transformer architectures such as AudioGen~\cite{audiogen}, CosyVoice~\cite{cosyvoice, cosyvoice2,cosyvoice3}, MAGNET~\cite{magnet}, and MusicGen~\cite{musicgen} demonstrate the effectiveness of discrete token modeling and autoregressive or masked generation for producing natural and coherent audio. In parallel, diffusion-based methods including DiffSound~\cite{diffsound}, AudioLDM~\cite{audioldm}, AudioLDM 2~\cite{audioldm2}, Tango~\cite{tango}, and Make-An-Audio~\cite{makeanaudio} leverage latent or discrete diffusion processes to achieve high-quality and controllable synthesis. More recently, hybrid designs incorporating DiT or flow-matching frameworks, such as TangoFlux~\cite{tangoflux}, FlashAudio~\cite{flashaudio}, and Audiobox~\cite{audiobox}, further improve fidelity, efficiency, and alignment with human preferences.

\begin{figure*}[t]
    \centering
    \includegraphics[width=1.0\linewidth]{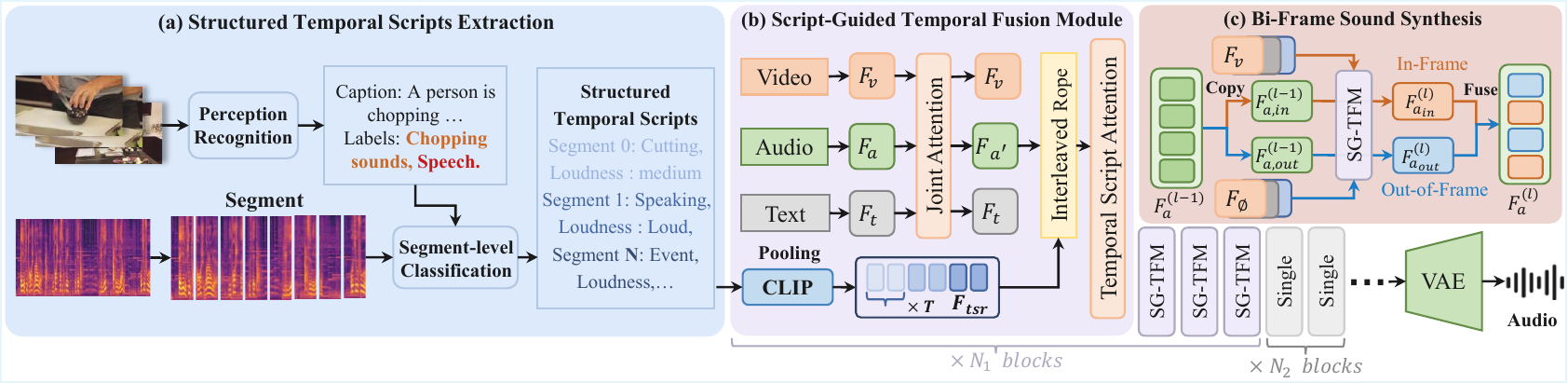}
    \caption{
        \textbf{Overview of our method.} (a) Extraction pipeline of segment-level \Feature~features. (b) Structure of the \Ada~module, where Temporal Script Attention introduces control signals. (c) \BiName, which leverages the controllability of our method in T2A and V2A to enable parallel rendering of in-frame and out-of-frame sounds. Fused block represents the single-modal transformer block in MMAudio.}
    \label{fig:overview}
    
\end{figure*}

\subsection{Video-to-Audio}
V2A generation has recently advanced through a variety of architectures that emphasize semantic alignment, temporal synchronization, and controllable sound synthesis. Early works such as FoleyCrafter~\cite{foleycrafter} and Diff-Foley~\cite{difffoley} leverage pre-trained text-to-audio or latent diffusion backbones with multimodal guidance to produce high-quality, video-synchronized audio. Models like Frieren~\cite{frieren} and VTA-LDM~\cite{vta-ldm} employ rectified flow or latent diffusion mechanisms with cross-modal fusion and auxiliary embeddings to enhance synchronization efficiency and semantic fidelity. Autoregressive frameworks such as V-AURA~\cite{V-AURA} further improve temporal precision through high-framerate visual feature extraction. Besides, MultiFoley~\cite{multifoley} and MMAudio~\cite{mmaudio} integrate text, audio, and video conditioning to enable flexible and high-fidelity sound generation; ThinkSound~\cite{thinksound} introduces chain-of-thought reasoning to support interactive and context-aware audio synthesis; and large multimodal diffusion-transformer frameworks such as HunyuanVideo-Foley~\cite{hunyuanfoley} and Kling-Foley~\cite{kling} extend this paradigm to unified text-video-to-audio modeling with enhanced synchronization, spatial awareness, and scalability.

However, existing V2A models rely on global text descriptions that provide only coarse semantic guidance, limiting their ability to capture fine-grained temporal details. Current open-source datasets further constrain control, as their captions are short and lack temporal annotations. In contrast, our method introduces fine-grained temporal control, enriching synthesis with detailed semantic cues and enabling more precise and creative audio generation.

\section{Method}
\label{sec:method}

\subsection{Overview}
\label{sec:overview}

\paragraph{Task Definition.} 
Given an input video $\mathcal{V}$ and its textual description $\mathcal{T}$, the conventional V2A task aims to generate an audio sequence $\mathcal{A}$ that is semantically consistent with $\mathcal{T}$ and temporally aligned with the visual events in $\mathcal{V}$. 
While the timing of sounds is often guided by visual cues, these cues are often incomplete or ambiguous in real-world videos, such as off-screen sounds, subtle motion in small regions, or occluded objects.
In such cases, providing the model with additional temporal and semantic cues beyond visual information can both \textit{compensate for missing visual features} and offer \emph{fine-grained temporal controllability.}
Specifically, as shown in Fig.~\ref{fig:show}, we aim to generate audio $\mathcal{A}$ that aligns with the video content where possible, while allowing users to supply temporal guidance for off-screen or visually subtle events.

\paragraph{Method Overview.}

As illustrated in Fig.~\ref{fig:overview}, \ModuleName~provides fine-grained temporal control built upon the pre-trained MMAudio generator~\cite{mmaudio}. First, as described in Section~\ref{sec:scripts}, we decompose the video into finer segments and introduce a segment-level pipeline to generate \FeatureName~(\Feature) features to provide \textit{richer temporal cues} for each segment. Next, in Section~\ref{sec:TFM}, we introduce \AdaName~(\Ada), an adapter that integrates these \Feature~features into the model in a temporally coherent manner. Finally, we utilize \ModuleName's \Feature~ controllability in both V2A and T2A tasks, propose \BiName~(\Bi) in Section~\ref{sec:dual_path}, enhance controllability in complex in-frame and out-of-frame scenarios.

\subsection{\FeatureName}
\label{sec:scripts}

In this section, we discuss how to construct features that \textit{provide additional temporal information beyond visual cues} while remaining easily \textit{interpretable by the model}.

\paragraph{Scripts Control in Finer Segment.} 

As described in Section~\ref{sec:overview}, we aim to incorporate more fine-grained temporal cues into the V2A process. A straightforward approach would be to use the global caption with complex temporal descriptions and explicit timestamps. However, in practice, users rarely provide precise timestamps, and detailed descriptions of complex sound events can be cumbersome and difficult for the model to interpret.
Inspired by recent advances in MIGC~\cite{migc, migc++},
we decompose the complex global caption into multiple fine-grained temporal captions (1s per segment), each corresponding to the semantic control of a specific audio segment. This structured temporal text, akin to a director’s script, allows users to specify the semantic content of each segment; we therefore refer to it as \FeatureName. STS effectively treated the entire audio generation process as multiple short segments, providing richer temporal cues for each segment, and providing short-range, segment-level global control.

\paragraph{\Feature~ Extraction.} 

As illustrated in Fig.~\ref{fig:overview} (a), we construct a fine-grained audio annotation pipeline that extracting \FeatureName, consisting of two main components:
\textbf{1) Perception and Recognition.}
We first leverage Qwen-Omni 7B~\cite{Qwen2.5-Omni} to generate content-aware captions for each audio clip, providing a global semantic description of the sounds. Based on this overall description, Qwen-Omni is then used to identify the types of sounds present in the current audio.
\textbf{2) Segment-level Classification.}
After forming a coarse understanding of the audio content, we then localize all potential audio events within each segment. As described in the previous section, we treat the \Feature~control as a global control at the segment-level. Therefore, we formulate the localization task as a segment-level binary classification problem to annotate the \Feature. Specifically, we divide the audio into 1-second segments and, based on the recognized audio events, employ a MLLM~\cite{Qwen2.5-Omni} to determine whether each audio event is present in the current segment. If an event is present, the model further provides intra-segment audio descriptions, loudness, and timbre. Using this pipeline, we collect the DirectorSound data set and the VGGSound-Director test set, which we describe in Section~\ref{sec:dataset}. We train our model only on the V2A DirectorSound dataset without any T2A dataset.

\subsection{\AdaName}
\label{sec:TFM}

After obtaining the \Feature, we need to incorporate it into the \ModuleName. Achieving an effective and harmonious integration poses the following challenges:
\textit{\textbf{C1}. How to extract \Feature~features that effectively capture temporal information;}
\textit{\textbf{C2}. How to fuse \Feature~features while preserve the model’s original generative capability;}
\textit{\textbf{C3}. How to achieve temporal alignment between \Feature~and audio Features.}

\paragraph{C1: \Feature~ Features.} 
We use $\mathcal{T}_{tsr}^i$ to denote the \FeatureName~within the $i$-th time segment. These textual caption can either be generated automatically by an MLLM or provided directly by the user.
As shown in Fig.~\ref{fig:overview}(b), following MMAudio’s modality processing strategy, we use a CLIP text encoder~\cite{clip} to extract textual features from each \Feature. Since these features contain many tokens, using too many temporal scripts would increase computational overhead and complicate temporal alignment. We thus use the pooling CLIP feature, yielding a compact temporal-semantic representation for the corresponding sound segment, denoted as $\mathbf{F}_{tsr}^i \in \mathbb{R}^{B \times 1 \times D}$:
\begin{gather}
    \mathbf{F}_{tsr}^i = \mathrm{Pool}\big(\mathrm{CLIP}(\mathcal{T}_{tsr}^i)\big), 
    \quad i = 1, 2, \dots, N,
\end{gather}
and all $\mathbf{F}^i_{tsr}$ are concatenated along sequence dimmension. Each segment-level \Feature~is replicated $T$ times to match the number of video tokens, preserving temporal consistency while providing sufficient semantic information. This yields the full \Feature~representation $\mathbf{F}_{tsr} \in \mathbb{R}^{B \times (T \times N) \times D}$:
\begin{gather}
    \mathbf{F}_{tsr} = [\mathbf{F}^1_{tsr}, \mathbf{F}^2_{tsr}, \dots, \mathbf{F}^N_{tsr}],
\end{gather}
where $N$ denotes the number of segments, $[\cdot]$ denotes the concatenation operation.
If a segment lacks an explicit description, we substitute it with an empty embedding.

\paragraph{C2: \AdaName.} 
Since training an audio generator from scratch requires massive computational resources, we aim to integrate \FeatureName~into the model while preserving the generative capability of the pretrained generator as much as possible. As illustrated in Fig.~\ref{fig:overview}(b), we design a \AdaName~(\Ada) that injects \Feature~features via an additional Temporal Script Attention (TSA), leaving the rest of MMAudio architecture unchanged. This independent attention layer allows the model to \textit{flexibly switch between traditional V2A and temporal script control simply by dropping \Ada.}
In the $l$-th block of MMAudio, given audio latents $\mathbf{F}_a^{(l)}$, video features $\mathbf{F}_v^{(l)}$, and global text features $\mathbf{F}_t^{(l)}$, we keep the joint attention unchanged:
\begin{equation}
\mathbf{F}_a^{(l)}, \mathbf{F}_v^{(l)}, \mathbf{F}_t^{(l)} = \mathrm{JointAttn}(\mathbf{F}_a^{(l-1)}, \mathbf{F}_v^{(l-1)}, \mathbf{F}_t^{(l-1)}),
\end{equation}
We then incorporate \Feature~features into the audio stream via TSA. Same as Joint Attention, audio features $\mathbf{F}_a^{(l)}$ are concatenated with \Feature~features $\mathbf{F}_{tsr}^{(l)}$, and unified self-attention fuses them into updated audio features $\mathbf{F}_{a'}^{(l)}$:
\begin{equation}
\mathbf{F}_{a'}^{(l)}, \mathbf{F}_{tsr}^{(l)} = \mathrm{TSA}(\mathbf{F}_a^{(l)}, \mathbf{F}_{tsr}^{(l-1)}).
\end{equation}

\renewcommand{\arraystretch}{1.25}

\begin{table*}[htbp]
\small
\centering
\begin{NiceTabular}{cccc@{\hspace{5pt}}ccc@{\hspace{5pt}}ccc@{\hspace{8pt}}c}
\toprule
\Block{2-1}{\textbf{Model}} & 
\multicolumn{3}{c}{\textbf{Counterfactual (C)}} & 
\multicolumn{3}{c}{\textbf{Temporal (T)}} & 
\multicolumn{3}{c}{\textbf{Overall (O)}} & 
\textbf{Quality (Q)} \\
\cmidrule(lr){2-4} \cmidrule(lr){5-7} \cmidrule(lr){8-10} \cmidrule(lr){11-11}
& \textbf{Precision}$\uparrow$ & \textbf{Recall}$\uparrow$ & \textbf{F1}$\uparrow$ & \textbf{Precision}$\uparrow$ & \textbf{Recall}$\uparrow$ & \textbf{F1}$\uparrow$ & \textbf{Precision}$\uparrow$ & \textbf{Recall}$\uparrow$ & \textbf{F1}$\uparrow$ & \textbf{FD$_{\text{VGG}}$$\downarrow$} \\
\midrule
MMAudio & 0.1775 & 0.2908 & 0.1825 & 0.2787 & 0.4591 & 0.2972 & 0.2281 & 0.3749 & 0.2378 & 8.55 \\
ThinkSound & 0.1141 & 0.2062 & 0.1208 & 0.2197 & 0.3042 & 0.2206 & 0.1669 & 0.2552 & 0.1707 & 7.17 \\
Video-Foley & 0.2009 & 0.3684 & 0.1976 & 0.2265 & 0.3674 & 0.2350 & 0.2137 & 0.3679 & 0.2163 & 8.03 \\
Hunyuan-Foley & 0.2241 & 0.3808 & 0.2331 & 0.2309 & 0.4265 & 0.2572 & 0.2275 & 0.4037 & 0.2451 & 7.51 \\
\toprule
\rowcolor{blue!10}
Ours & \textbf{0.5226} & \textbf{0.6101} & \textbf{0.5284} & \textbf{0.4129} & \textbf{0.5823} & \textbf{0.4354} & \textbf{0.4677} & \textbf{0.5962} & \textbf{0.4819} & \textbf{6.19} \\
\bottomrule
\end{NiceTabular}
\caption{Evaluation on DirectorBench. \textbf{C} denotes Counterfactual control, \textbf{T} denotes Temporal control, \textbf{O} denotes Overall performance, and FD$_{\text{VGG}}$ (lower is better) measures visual quality.}
\label{tab:controllability_metrics}
\vspace{-10pt}
\end{table*}

\paragraph{C3: Interleaved RoPE for Temporal Alignment.} 
Since \Feature~features have temporal characteristics, we draw inspiration from HunyuanVideo-Foley~\cite{hunyuanfoley} and introduce an interleaved RoPE encoding mechanism to enhance temporal alignment during feature fusion. Specifically, we first upsample \Feature~tokens to match the temporal resolution of audio features, then interleave audio and \Feature~features along the temporal dimension:
\begin{equation}
\mathbf{F}_{int} = \mathrm{Interleave}(\mathbf{F}_a^{(l)}, \mathrm{Up}(\mathbf{F}_{tsr}^{(l-1)})),
\end{equation}
where $\mathrm{Up}(\cdot)$ is the upsampling operation, $\mathrm{Interleave}(\cdot)$ interleaves neighboring tokens, and $\mathbf{F}_{int}$ denotes interleaved audio-\Feature~features. We then apply RoPE to $\mathbf{F}_{int}$, separate the features back into audio and \Feature~features, downsample them to their original temporal lengths, and perform TSA:
\begin{equation}
\mathbf{F}_{rope} = \mathrm{RoPE}(\mathbf{F}_{int}),
\end{equation}
\begin{equation}
\begin{aligned}
\mathbf{F}^{(l)}_{a,rope}, \mathbf{F}^{(l-1)}_{tsr,rope} &= \mathrm{UnInterleave}(\mathbf{F}_{rope}), \\
\mathbf{F}_{a'}^{(l)},\mathbf{F}_{tsr}^{(l)}  &= \mathrm{TSA}(\mathbf{F}^{(l)}_{a,rope}, \mathbf{F}^{(l-1)}_{tsr,rope}),
\end{aligned}
\end{equation}
Interleaved RoPE assigns correlated positional indices to temporally adjacent features, enabling temporally coherent \Feature~fusion.

\renewcommand{\arraystretch}{1.25}

\begin{table*}[t]
\small
\centering
\begin{NiceTabular}{@{\hspace{2pt}}>{\centering\arraybackslash}p{2.2cm}
                    @{\hspace{2pt}}ccccccccc@{\hspace{2pt}}}
\toprule
 \Block{2-1}{\textbf{Model}}  & \multicolumn{5}{c}{{\textbf{Distribution matching}}} & {\textbf{Audio quality}} & {\textbf{Semantic align}} & {\textbf{Temporal align}} \\
\cmidrule(lr){2-6} \cmidrule(lr){7-7} \cmidrule(lr){8-8} \cmidrule(lr){9-9}
& \textbf{FD$_{\text{VGG}}$} $\downarrow$ & \textbf{FD$_{\text{PANN}}$} $\downarrow$ & \textbf{FD$_{\text{PaSST}}$}$\downarrow$ & \textbf{KL$_{\text{PANN}}$}$\downarrow$ & \textbf{KL$_{\text{PaSST}}$}$\downarrow$ & \textbf{ISC$_{\text{PANN}}$}$\uparrow$ & \textbf{IB}$\uparrow$ & \textbf{DeSync}$\downarrow$ \\
\midrule
\rowcolor{defaultColor}
GT & 0.00 & 0.00 & 0.00 & 0.00 & 0.00 & 12.73 & 0.33 & 0.625 \\
\toprule
MMAudio & 1.45 & \textbf{8.23} & \textbf{92.80} & 1.67 & 1.47 & 14.38 & 0.32 & 0.439 \\
Hunyuan-Foley & 2.39 & 14.10 & 114.09 & 2.03 & 1.81 & 12.74 & 0.31 & 0.543 \\
Thinksound & 2.75 & 17.32 & 174.80 & 2.11 & 1.92 & 8.13 & 0.22 & 0.560 \\
\bottomrule
\rowcolor{blue!10}
Ours(w/o \Feature) & 1.27 & 8.58 & 94.15 & 1.65 & 1.43 & 13.81 & 0.32 & 0.438 \\
\rowcolor{blue!10}
Ours & \textbf{1.17} & 8.60 & 94.27 & \textbf{1.42} & \textbf{1.23} & \textbf{14.84} & \textbf{0.33} & \textbf{0.432} \\
\toprule
\end{NiceTabular}
\caption{Comparison of V2A models on distribution matching, audio quality, semantic alignment, and temporal alignment metrics. Lower values ($\downarrow$) indicate better performance for FD, KL, and DeSync; higher values ($\uparrow$) indicate better performance for ISC and IB-score. We also report inference w/o \Feature~tokens, demonstrates that our method can flexibly switch between V2A and temporally controlled modes.}
\label{tab:v2a_metrics}
\vspace{-10pt}
\end{table*}

\subsection{\BiName}
\label{sec:dual_path}

\paragraph{Challenge: Complex In-frame/Out-of-frame Scenarios.}
Introducing \Feature~features provides temporal control over audio generation. However, users may encounter more complex scenarios involving multi-event in-frame and out-of-frame sounds, or even counterfactual combinations. For instance, a user might want to generate a sound where a person suddenly utters a humorous or off-gender voice while speaking, or a dog bark unexpectedly occurs amid a chorus of birds. 
In such cases, visual cues can dominate, causing the model to ignore text-specified attributes unrelated to the visuals and lose control.
Therefore, we aim to explore methods to further improve the framework’s controllability to perform flexible in-frame/out-of-frame synthesis.

\paragraph{Bi-Frame Sound Synthesis.}  
Building upon pretrained MMAudio’s multimodal generative capability, we find that training \ModuleName~on the V2A dataset (DirectorSound) \textit{without T2A data} preserves temporal script control in T2A generation. Motivated by this, we propose the \BiName~(\Bi), combining MMAudio’s audio generation with our module’s temporal control to achieve \textbf{attribute disentanglement} for in-frame and out-of-frame sounds, enhancing both \textbf{controllability} and \textbf{semantic expressiveness}.  

As shown in Fig.~\ref{fig:overview}(c), generation is split into:  
1) \textbf{In-frame sounds}: events visible in the video;  
2) \textbf{Out-of-frame sounds}: off-screen or narrative events.  
Given audio latent $\mathbf{F}_a$, video features $\mathbf{F}_v$, text control $\mathbf{F}_t$, and \Feature~condition $\mathbf{F}_{tsr}$, we duplicate $\mathbf{F}_a$ into two streams and process each block $l$ in parallel:  

\begin{equation}
\begin{aligned}
    \mathbf{F}_{a,in}^{(l)},... &= \mathrm{Block}(\mathbf{F}_a^{(l-1)}, \mathbf{F}_v^{(l-1)}, \mathbf{F}_t^{(l-1)}, \mathbf{F}_{tsr}^{(l-1)}), \\
    \mathbf{F}_{a,out}^{(l)},... &= \mathrm{Block}(\mathbf{F}_a^{(l-1)}, \mathbf{F}_v^{\varnothing}, \mathbf{F}_t^{(l-1)}, \mathbf{F}_{tsr}^{(l-1)}),
\end{aligned}
\end{equation}  

In-frame audio $\mathbf{F}_{a,in}^{(l)}$ is conditioned on video, text, and \Feature; out-of-frame audio $\mathbf{F}_{a,out}^{(l)}$ uses a learnable null visual embedding, with text and \Feature~as control, while Synchformer features remain unchanged.  
Within each \AdaName, the two streams are fused according to temporal segments:  
\begin{equation}
    \mathbf{F}_a^{(l)} = \mathrm{Fuse}(\mathbf{F}_{a,in}^{(l)}, \mathbf{F}_{a,out}^{(l)}),
\end{equation}  
where $\mathrm{Fuse}$ concatenates in-frame and out-of-frame segments in temporal order. This bi-frame design enables joint rendering of on- and off-screen sounds, preserves temporal coherence, and enhances controllability.

\section{DirectorSound and Benchmarks}
\label{sec:dataset}

Following the pipeline described in Section~\ref{sec:scripts}, we collected and constructed a set of training data, and built a corresponding benchmark to evaluate the model’s generation quality and control capabilities.

\subsection{Data Source}
We construct the DirectorSound dataset from multiple sources.
First, we utilize several existing open-source V2A datasets, including VGGSound~\cite{vggsound} and AudioCaps~\cite{audiocaps}.
VGGSound is a large-scale audio-visual correspondence dataset containing over 200,000 video clips collected from YouTube, covering more than 310 sound classes and totaling about 550 hours of data. 
AudioCaps, on the other hand, is built upon the AudioSet corpus and provides human-written natural language descriptions for 10-second audio clips.
In addition, we also incorporate a portion of our in-house V2A dataset to further enrich the diversity and coverage of DirectorSound.

\subsection{DirectorBench and VGGSound-Director}
\label{sec:bench}

To evaluate our method, we construct two benchmark: DirectorBench and VGGSound-Director.
\textbf{DirectorBench} focuses on evaluating the model’s \textbf{\textit{Controllability}}. We collected 100 videos from the VGGSound test set and Pexels, and manually annotated the sound-emitting segments in two types: \textit{1) Temporal Control}. Involves out-of-frame sounds or duration control, such as specifying when a sound should occur or enforcing silence in certain audio segments. \textit{2) Counterfactual Control}. Introduces counterfactual sound attributes, generating audio that contradicts the video content in specific actions, or allowing flexible combinations of conventional and counterfactual sounds. For each video, we annotated approximately 1–4 control templates per sound type, resulting in roughly 400 test samples in total.
During evaluation, we employ the audio grounding tool~\cite{grounding} to locate target vocalization segments and compute the intersection-over-union (IoU) with the annotated time segments. Segments exceeding a predefined threshold are considered correctly triggered, from which we calculate Precision, Recall, and F1 score. 
Besides, we also want to evaluate the audio \textbf{\textit{Quality}} we generated, so we design another benchmark named \textbf{VGGSound-Director}. VGGSound-Director is a subset of VGGSound containing 2.2K videos, annotated using the temporal labeling procedure described in Section~\ref{sec:scripts}. These annotations are based on real audio, allowing us to evaluate the \textit{Quality} of controllable synthesized audio. \textbf{Additional information can be found in the supplementary materials.}

\section{Experiments}
\label{sec:exp}

\subsection{Details}

\noindent \textbf{Impletmetation.} All experiments are conducted using MMAudio-medium with full model training. We set the learning rate to 2e-5 and employ a batch size of 16. During training, \Feature~features are randomly dropped with a probability of 0.1 (replaced with empty text features) to enable Classifier-free Guidance(CFG). A cosine learning rate scheduler is used, and the model is trained for 1.2 million iterations. Training is performed on 8 × 40GB A800 GPUs, with a total training time of approximately 3 days.
At inference, we adopt MMAudio’s default configuration, run 25 inference steps, and set the CFG scale to 4.5.

\noindent \textbf{Baselines.} We compare our method with several state-of-the-art V2A models, including MMAudio, HunyuanVideo-Foley, and ThinkSound. For controllability comparisons, in addition to these methods, we also include Video-Foley~\cite{lee2025video}, which uses video-extracted RMS signals to control the generation process. On VGGSound-Director, all models use the original VGGSound captions, while on DirectorBench, we manually annotate temporal captions, including the timing of vocalizations. All methods are evaluated using their default parameter configurations.

\renewcommand{\arraystretch}{1.25} 

\begin{table}[t]
\small
\centering

\begin{NiceTabular}{c cccc}
\toprule
\Block{2-1}{\textbf{ID}} & \Block{2-1}{\textbf{Model}} &  \multicolumn{3}{c}{\textbf{Overall (O)}} \\
\cmidrule(lr){3-5}
& & \textbf{Precision}$\uparrow$ & \textbf{Recall}$\uparrow$ & \textbf{F1}$\uparrow$ \\
\midrule
\rowcolor{green!10}
\ding{192} & Base & 0.1448 & 0.1963 & 0.1311  \\
\rowcolor{green!10}
\ding{193} & + \Feature & 0.4102 & 0.5432 & 0.4252 \\
\rowcolor{green!10}
\ding{194} & + Interleaved Rope & 0.4209 & 0.5582 & 0.4389 \\
\rowcolor{green!10}
\ding{195} & + \Bi & \textbf{0.4677} & \textbf{0.5962} & \textbf{0.4819} \\
\toprule
\rowcolor{blue!10}
\ding{196} & $\text{w/o \Bi}_{\text{(sub)}}$ & 0.3928 & 0.5373 & 0.4178 \\
\rowcolor{blue!10}
\ding{197} & $\text{w \Bi}_{\text{(sub)}}$ & \textbf{0.4449} & \textbf{0.5701} & \textbf{0.4613} \\
\bottomrule
\end{NiceTabular}
\caption{Ablation on DirectorBench. \textbf{O} denotes Overall performance. Green rows report testing on the full benchmark, blue rows report test on the reannotated subset.}
\label{tab:ablation}
\vspace{-5pt}
\end{table}

\noindent \textbf{Metrics.}
On VGGSound-Director, we evaluate our generated audio using five complementary metrics: \textbf{1) Distribution Matching}: We measure how closely the generated audio matches the ground-truth distribution using Fréchet Distance (FD) and KL divergence (KL). FD is computed with embeddings from PaSST~\cite{passt}, PANNs~\cite{panns}, and VGGish~\cite{vggish}, while KL uses PaSST and PANNs as classifiers. \textbf{2) Audio Quality}: To assess perceptual quality without reference audio, we use the Inception Score (ISC), with PANNs as the classifier.
\textbf{3) Semantic Alignment}: We measure the semantic consistency between the input video and generated audio using ImageBind~\cite{imagebind}, reporting the average cosine similarity as the IB-score.
\textbf{4) Temporal Alignment}: Audio-visual synchrony is evaluated using DeSync~\cite{synchformer}, predicted by Synchformer, which estimates the timing misalignment between audio and video. 
On DirectorBench, we divide controllability into temporal control and counterfactual control, and evaluate metrics for each case as well as overall performance. As described in Section~\ref{sec:bench}, we use grounded audio event timestamps to match with the target time segments and compute the IoU. Vocalizations exceeding the threshold are considered correct, from which we calculate Precision, Recall, and F1 score.

\subsection{Result}

\begin{figure*}[t]
    \centering
    \includegraphics[width=0.9\linewidth]{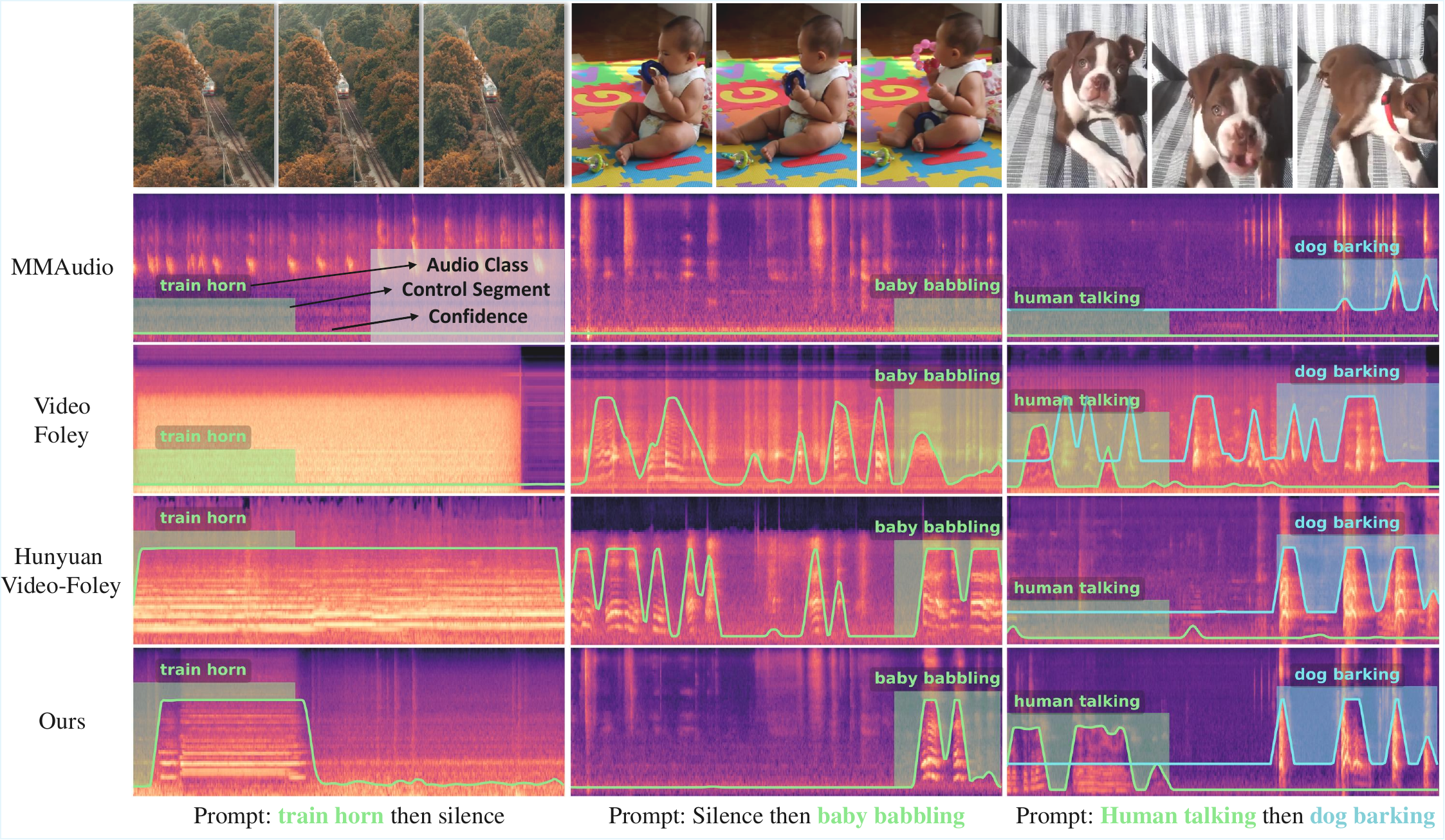}
    \caption{
        \textbf{Qualitative comparison.} We visualize the spectrograms of generated audio (by prior works and our method) and the ground-truth. The curves in the figure represent the recognition confidence for the corresponding sound categories, while the rectangular regions along the curves indicate the time intervals where our fine-grained scripts provide control.}
    \label{fig:result}
    \vspace{-10pt}
\end{figure*}

\paragraph{DirectorBench.} 
From Tab.~\ref{tab:controllability_metrics}, we conclude that our approach \textbf{enhances temporal control capability while preserving audio quality}. Our method achieves substantial improvements over prior approaches in both counterfactual audio control and explicit control over the presence or silence of specific sound segments. On the counterfactual subset, the F1 score is boosted from 0.2331 to 0.5284, and on the temporal subset, it rises from 0.2572 to 0.4354. Furthermore, we attain the lowest $\text{FD}_\text{VGG}$ score among all methods, demonstrating that incorporating \textit{Feature} strengthens visual--temporal control without compromising generation quality.

\noindent \textbf{VGGSound-Director.} 
From Tab.~\ref{tab:v2a_metrics}, we can draw the following conclusions: \textbf{1) Our method preserves the baseline generation capability of MMAudio.} With \Feature~(last row in table), the audio generated also does not show any significant degradation in FD$_{\text{PANN}}$ or FD$_{\text{PaSST}}$, with slight improvement in IB-score and Desync. \textbf{2) Introducing \Feature~does not noticeably harm the quality of the generated audio.} Incorporating \Feature~brings the generated audio closer to real recordings. Our method reduces FD$_{\text{VGG}}$ from 1.45 to 1.17, and KL$_{\text{PANN}}$ and KL$_{\text{PaSST}}$ from 1.67 and 1.47 to 1.42 and 1.23, respectively. ISC rises from 14.38 to 14.84. 
\textbf{3) We provide a flexible control framework} that allows users to \textit{seamlessly switch between traditional V2A generation and temporally controlled modes}. We simply removed \Ada~(Ours w/o \Feature), the generated audio shows no significant degradation in terms of distribution, quality, or semantic consistency compared with MMAudio.

\begin{table}[t]
\setlength{\abovecaptionskip}{0pt}
\setlength{\belowcaptionskip}{0pt}
\footnotesize
\centering
\renewcommand{\arraystretch}{0.75}

\begin{subtable}[t]{0.49\columnwidth}
\centering
\caption{User Study}
\label{tab:user_study}
\begin{NiceTabular}{c@{\hspace{2pt}}c@{\hspace{2pt}}c@{\hspace{2pt}}c}
\toprule
\textbf{Method} & \textbf{Qua.}$\uparrow$ & \textbf{Ctrl.}$\uparrow$ & \textbf{Align.}$\uparrow$\\
\midrule
MMAudio & 3.03 & 2.67 & 2.60   \\
HunyuanFoley    & 3.07 & 2.93 & 3.10 \\
Ours    & \textbf{4.27} & \textbf{4.50} & \textbf{4.53} \\
\bottomrule
\end{NiceTabular}
\end{subtable}
\hfill
\begin{subtable}[t]{0.49\columnwidth}
\centering
\caption{STS duration}
\label{tab:STS duration}
\begin{NiceTabular}{c c}
\toprule
\textbf{Duration} & \textbf{F1(O)}$\uparrow$ \\
\midrule
1s         & 0.4819 \\
0.5s      & \textbf{0.5197} \\
2s     & 0.4646 \\
\bottomrule
\end{NiceTabular}
\end{subtable}
\caption{User Study and STS duration experiments}
\label{tab:rebuttal}
\vspace{-10pt}
\end{table}

\noindent \textbf{Qualitative Results.}
We present qualitative comparisons with several baselines in Fig.~\ref{fig:result}. In addition to Mel-spectrograms of the generated audio, we visualize grounding model confidence curves for different sound events and mark user-controlled segments with rectangles. For example, in the leftmost case, we control the train horn sound within the first three seconds. We draw three main observations:
\textbf{1) Enhanced controllability.}
Without clear visual cues, previous methods struggle to control the exact timing of sound generation. HunyuanVideo-Foley often generates sound across the entire segment (yielding high recall), while others produce unrecognizable or unconstrained sounds. Our method reliably synthesizes recognizable audio precisely within the user-controlled intervals.
\textbf{2) Effective collaboration with visual features.}
In the rightmost example, providing an \Feature~description aligned with the visual content (dog barking, 6–8s) allows our model to generate audio matching the rhythm of HunyuanVideo-Foley. This shows that \Feature~provides fine-grained temporal guidance without overriding video-based cues, enabling global control within segments while preserving the model’s understanding of visual features.
\textbf{3) Flexible attribute control.}
Our method can also synthesize off-screen sounds, e.g., human speech at the beginning, enabling richer storytelling beyond the visible scene. \textbf{Additional results are provided in the supplementary materials.}

\subsection{Analysis}

\noindent \textbf{User Study.} 
We invited 30 users to rate samples from FoleyDirector, MMAudio and HunyuanVideo-Foley on Quality (Qua.), Controllability (Ctrl.) and Temporal Alignment (Align.) (1–5 scale) in Tab.\ref{tab:user_study}. It demonstrates consistency between objective and subjective evaluations, confirming FoleyDirector's superior controllability.

\noindent \textbf{STS Segment Duration.}
We directly infer on different STS durations (Tab.\ref{tab:STS duration}) using a model trained \textbf{only on 1s segment data} (e.g., splitting 8 STS tokens of each 1s segment into two 0.5s segments with 4 tokens each). 
\textbf{1)} Models trained on 1s segments still enable temporal control over other STS durations. 
\textbf{2)} Controllability correlates with STS duration: longer segments have fewer temporal details and lower F1 scores, while shorter segments yield more temporal details and higher F1 scores. 
\textbf{3)} STS duration selection involves a \textbf{trade-off} between: \textit{a)} Controllability (shorter durations add more temporal details; 1s suffices for strong performance); \textit{b)} Usability (shorter durations require more user-provided scripts, e.g., 16 STS for 0.5s segments); \textit{c)} Annotation Accuracy (overly short audio clips lack contextual information, increasing data annotation error rates).

\subsection{Ablation}
\noindent \textbf{\Feature.}
As shown in the Tab.~\ref{tab:ablation} ID \ding{172} and ID \ding{173}, +\Feature~incorporates \Feature~using an additional Temporal Script Attention without Interleaved RoPE. We observe that introducing the \Feature~significantly enhances the model’s controllability on the DirectorBench. Precision increases from 0.1448 to 0.4102, Recall rises from 0.1963 to 0.5432, and F1 score improves from 0.1311 to 0.4252.

\noindent \textbf{Interleaved Rope.} By comparing ID \ding{173} and ID \ding{174} in the Tab.~\ref{tab:ablation}, we can observe that introducing the Interleaved RoPE improves temporal alignment, thereby enhancing the overall model performance.

\noindent \textbf{\Bi.} On the DirectorBench, we evaluate performance both with (w/ \Bi) and without (w/o \Bi) the Bi-Frame Sound Synthesis Framework.
To enable the evaluation, we further annotate a subset of complex cases to distinguish between in-frame and out-of-frame sounds. We report results on both the entire benchmark (\ding{174}, \ding{175}) and the annotated subset (\ding{176}, \ding{177}) in Tab.~\ref{tab:ablation}.
We observe that:
\textbf{1)} The annotated subset is inherently more difficult, with notably lower precision compared to the overall benchmark;
\textbf{2)} The proposed framework effectively mitigates this difficulty, improving F1 from 0.4389 to 0.4819, and even surpassing the overall baseline performance on the challenging subset.
\section{Conclusion}
\label{sec:conclusion}

We presented \ModuleName, empowering users to act as foley directors who can control when and how sounds emerge.
We enhance the pretrained MMAudio with additional temporal cues beyond visual features, decomposing videos into finer segments with \FeatureName~for each. We also propose a \AdaName to introduce these temporal features, allow user to seamlessly switch between traditional V2A generation and temporally controll by disabling \Ada. We also propose \BiName to enable more flexible in and out of frame audio generation.
We construct a annotation pipeline and DirectorSound dataset. Besides, we propose DirectorBench and VGGSound-Director to evaluate the controllability and quality.
Extensive experiments highlight improvements in precision and F1-score, validating the effectiveness of our method in advancing controllable audio generation.

\noindent \textbf{Acknowledgements}. This work was supported by the Fundamental Research Funds for the Central Universities (226-2025-00080), the National Natural Science Foundation of China (62293554)
and in part by the Natural Science Foundation of Zhejiang Province (LDT23F02023F02)
{
    \small
    \bibliographystyle{ieeenat_fullname}
    \bibliography{main}
}

\clearpage
\appendix 
\setcounter{page}{1}

\maketitlesupplementary

\begin{figure}[ht]
    \centering
    \includegraphics[width=1.0\linewidth]{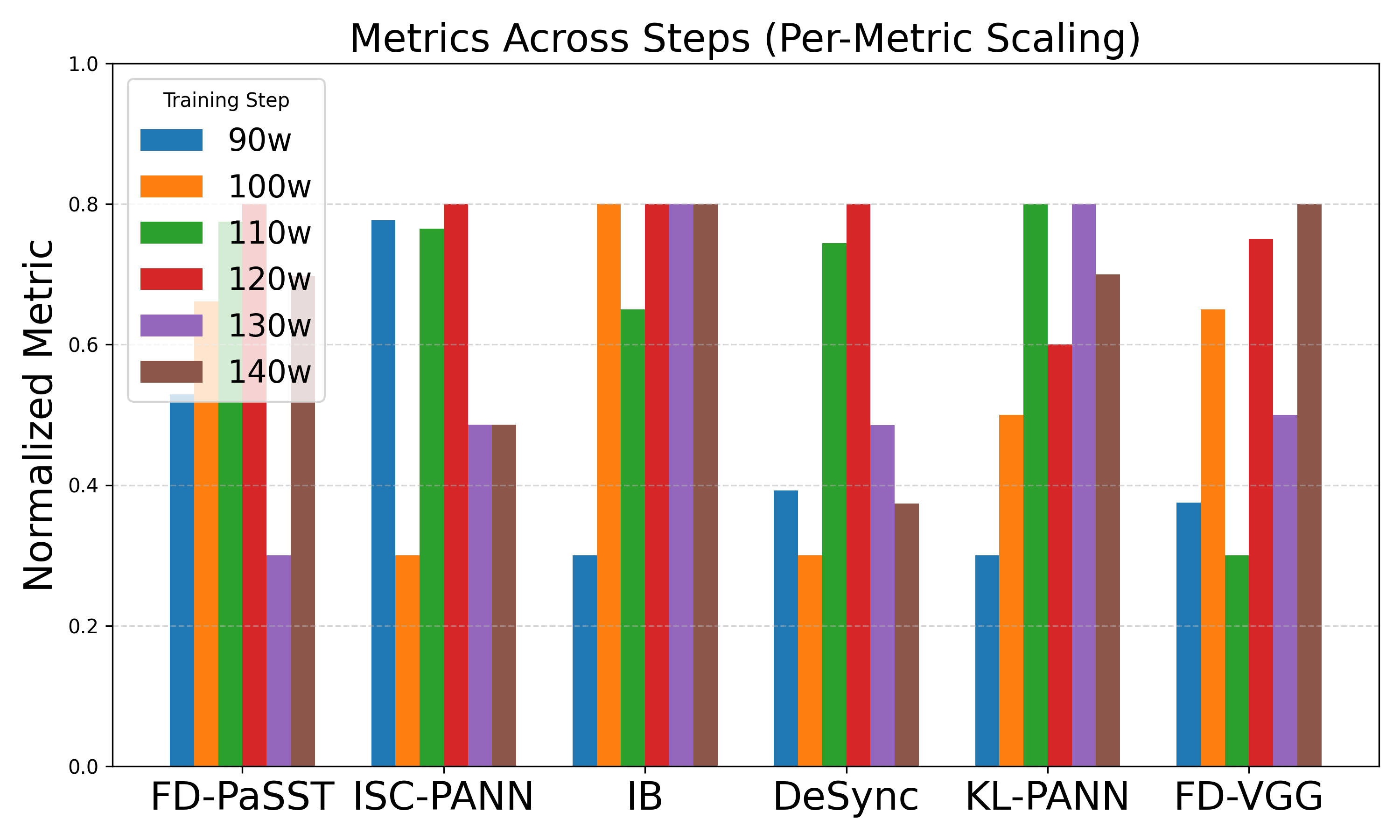}
    \caption{
        \textbf{Ablation on training iterations.} We show the model metrics after training for different numbers of iterations. The metrics are normalized to the range of 0.3–0.8, with higher values indicating better performance.}
    \label{fig:ablation_iters}
\end{figure}

\renewcommand{\arraystretch}{1.25}

\section{Model Architecture}
\label{sec:architecture}

\noindent \textbf{DiT Architecture.}
Our model architecture is designed following the framework of MMAudio, consisting of stacked multi-stream and single-stream DiT blocks. The video input (8 frames per second) is encoded by CLIP to extract visual features, while the global text description is processed by CLIP’s text encoder to obtain textual representations. The audio input is encoded and compressed using the VAE pretrained in MMAudio. All these modality-specific features are projected through their respective projectors and fed into the multi-stream DiT blocks. Additionally, the pooled global tokens from the video and text encoders, together with timestep and Synchformer features, are used to construct a temporally-aware global embedding, which conditions the adaptive LayerNorms within each DiT block.

In the DiT block, we further improve upon MMAudio’s original design. After adaptive LayerNorm, temporal RoPE encodings are added to both video and audio features, and joint attention is performed across the three modalities (video, text, and audio). Following this joint attention, we introduce an additional SG-TFM module, which takes STS features as extra input and performs Temporal Script Attention, a type of joint attention with the audio features to inject fine-grained temporal-semantic information. After passing through four joint blocks, the fused audio features, which integrate visual, textual, and temporal cues, are processed by eight single-stream DiT blocks (as in MMAudio) to produce the final target flow prediction.

When users wish to render complex soundscapes with fine-grained control, we introduce an Bi-Frame Sound Synthesis Framework. In each joint block, the audio latent is copied into in-frame and out-of-frame branches. During rendering, we disable the influence of visual features on the out-of-frame  branch, allowing it to be modulated solely by textual and temporal-semantic cues, while maintaining conditioning from the globally temporal Synchformer embedding. The in-frame branch, in contrast, follows the standard rendering process. The latent outputs from both the in-frame and out-of-frame SG-TFM modules are then segmented and reorganized according to the control intervals of the desired sound events, enabling synchronized yet independently controllable generation of in-frame and out-of-frame audio.

\begin{figure*}[ht]
    \centering
    \includegraphics[width=1.0\linewidth]{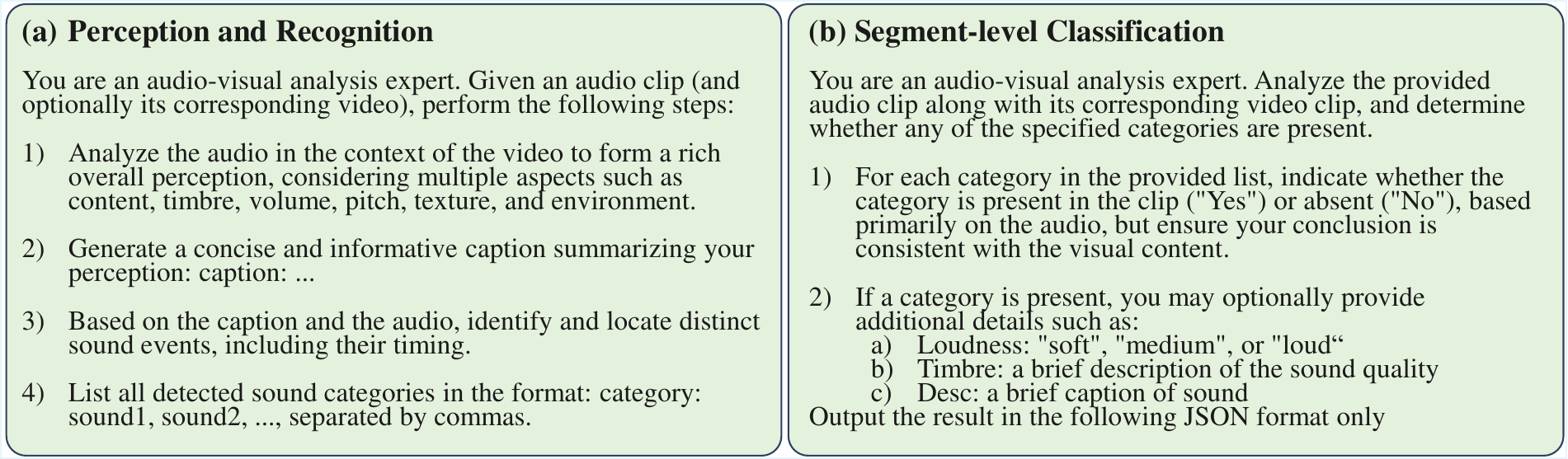}
    \caption{
        \textbf{System Prompt.} The system prompt we used in annotation pipeline}
    \label{fig:system_prompt}
\end{figure*}

\noindent \textbf{VAE and vocoder.}
In our framework, we adopt the audio representation and backbone modules directly from MMAudio, including its pretrained VAE encoder–decoder and BigVGAN vocoder.
Following MMAudio, the audio waveform is first transformed into mel spectrograms by applying a short-time Fourier transform (STFT) and extracting the magnitude component. The resulting spectrograms are then encoded into latent representations using a pretrained variational autoencoder (VAE). During inference, generated latents are decoded back into mel spectrograms by the same VAE, which are subsequently converted into audio waveforms using the pretrained vocoder.
The VAE follows the 1D convolutional network design of Make-An-Audio 2, employing a downsampling factor of 2 and trained with reconstruction, adversarial, and Kullback–Leibler (KL) divergence objectives. MMAudio applies the magnitude-preserving network design introduced in EDM2, replacing standard convolutional, normalization, addition, and concatenation layers with magnitude-preserving counterparts. This modification effectively stabilizes latent magnitudes without noticeably affecting reconstruction performance.
For vocoder components, we use the 44.1 kHz version BigVGAN-v2.

\renewcommand{\arraystretch}{1.25}

\begin{table*}[t]
\small
\centering
\begin{NiceTabular}{@{\hspace{1pt}}>{\centering\arraybackslash}p{2.2cm}
                    @{\hspace{1pt}}cccccccccc@{\hspace{1pt}}}
\toprule
 \Block{2-1}{\textbf{Model}}  & & \multicolumn{5}{c}{{\textbf{Distribution matching}}} & {\textbf{Quality}} & {\textbf{Semantic}} & {\textbf{Temporal}} \\
\cmidrule(lr){2-2} \cmidrule(lr){3-7} \cmidrule(lr){8-8} \cmidrule(lr){9-9} \cmidrule(lr){10-10}
& Params & \textbf{FD$_{\text{VGG}}$} $\downarrow$ & \textbf{FD$_{\text{PANN}}$} $\downarrow$ & \textbf{FD$_{\text{PaSST}}$}$\downarrow$ & \textbf{KL$_{\text{PANN}}$}$\downarrow$ & \textbf{KL$_{\text{PaSST}}$}$\downarrow$ & \textbf{ISC$_{\text{PANN}}$}$\uparrow$ & \textbf{IB}$\uparrow$ & \textbf{DeSync}$\downarrow$ \\
\midrule
Medium & 621M & 1.45 & 8.23 & 92.80 & 1.67 & 1.47 & 14.38 & 0.32 & 0.439 \\
Large & 1.03B & 1.47 & 8.22 & 93.77 & 1.66 & 1.46 & 14.41 & 0.32 & 0.456 \\
\toprule
\end{NiceTabular}
\caption{Comparison of MMAudio models of different sizes on VGGSound-Director.All the model are 44KHZ.}
\label{tab:model_size}
\end{table*}

\section{Pipeline Details}
\label{sec:pipeline details}

\begin{figure*}[ht]
    \centering
    \includegraphics[width=1.0\linewidth]{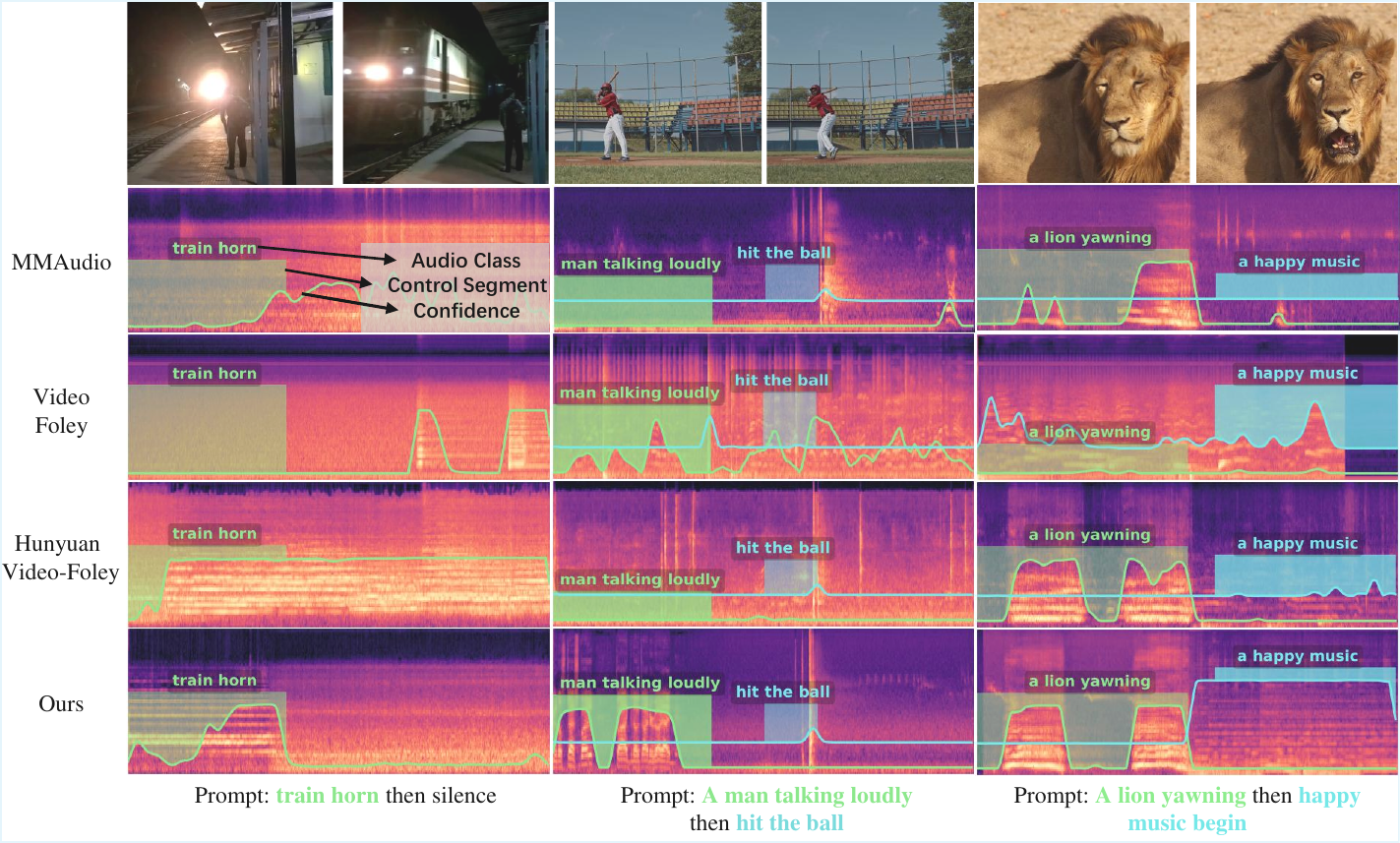}
    \caption{
        \textbf{Visual Results in DirectorBench.} We present several results from DirectorBench.}
    \label{fig:ablation_result_directorbench}
\end{figure*}

We show the details of our data annotation pipeline, including Perception and Recognition, segment-level Classification.

\noindent \textbf{Perception and Recognition.}
We use system instructions to prompt Qwen-Omni to generate a caption for the given audio, enabling it to form a holistic understanding of the audio content. Based on this overall comprehension, we further instruct Qwen-Omni to identify the types of sounds present in the audio and return their corresponding category names. Through this process, we obtain a fundamental understanding of the audio and discover the fine-grained sound events it contains. The system prompt we use is shown in Fig.~\ref{fig:system_prompt} (a).

\noindent \textbf{Segment-level Classification.}
We further perform \textbf{segment-level classification} to obtain fine-grained audio labels at the temporal segment level.
Directly prompting Qwen-Omni to localize audio events yields unsatisfactory precision.
Therefore, we reformulate the localization problem into a \textit{binary classification task}.
Specifically, we pre-segment each audio clip (and its corresponding video) into fixed-length intervals, and instruct Qwen-Omni to determine whether each predefined sound category is present within a given segment.
If a sound is detected, Qwen-Omni is further asked to describe its content and timbre characteristics, thereby producing fine-grained segment-level text annotations.
The system prompt used for this stage is illustrated in Fig.~\ref{fig:system_prompt} (b).

\section{More Analysis}
\label{sec:architecture}

\noindent \textbf{The difference between other works.} 
1) The concept of \textbf{temporal control} refers to controlling the generation of both \textit{on-screen} and \textit{off-screen} sounds \textit{within specific time periods} in our paper. Given a car video $V_{car}$, we can control the car horn  at arbitrary time periods, which is an active control capability that \textbf{the other two works fail to achieve}. 2) \textbf{VTA-SAM}~\cite{hu2024video} only extracts frame-level object features (not event features) via SAM, it can only extract the information indicating the probability of a car being present in each frame of $V_{car}$, unable to provide information for objects that are off-screen, and such object-level feature, rather than event-based feature, also fail to enable the control of the car horn at any arbitrary time.
3) \textbf{Video-Foley}~\cite{lee2025video} \textit{relies entirely on Videos} to extract RMS signals (thus cannot control off-screen sounds). The RMS signal only captures temporal intensity, lacks semantic-temporal information for multiple events to achieve fine-grained semantic-temporal control.

\noindent
\noindent \textbf{Details on inference latency.}
We will add test latency details in the revised version. We tested the average inference time on several 8-second video with MMAudio (\textbf{3.30s}) and our method (\textbf{3.76s}) on A800. It shows that we \textbf{introduces only about 12\% additional inference overhead}, but achieves a \textbf{significant 102.6\% performance improvement} (from 0.2378 to 0.4819) on Overall F1 score.

\section{The Composition of STS}
\label{sec:sts}

Through the above pipeline, we can obtain the audio events and their associated attribute descriptions (Loudness and Timbre) for each segment. We combine these descriptions using a simple template (f"\{Desc\}, \{Loudness\}, \{Timbre\}") to generate the Structured Temporal Scripts for each segment.

\section{Benchmark Details}
\label{sec:results}

In this section, we present the details of the benchmark we constructed. The benchmarks used in our evaluation include \textbf{DirectorBench} and \textbf{VGGSound-Director}.

\noindent \textbf{DirectorBench.}
DirectorBench primarily evaluates a model’s controllability. To build this benchmark, we collected videos from the VGGSound test set and the Pexels website, and manually annotated various control instructions. The annotation process is as follows:
\begin{enumerate}
    \item \textbf{Anchoring existing sound events.} All annotations are based on sound events that can be visually identified, so we first locate visible sound events in each video.
    \item \textbf{Temporal control.} This category examines the model’s ability to control when sounds occur and includes:
    \begin{enumerate}
        \item Allowing certain sound events to occur normally.
        \item Enforcing silence for specific sound events.
        \item For sound events with uncertain onset (e.g., distant fireworks or thunder), assigning a randomly selected segment as the sounding period.
    \end{enumerate}
    \item \textbf{Counterfactual control.} This category evaluates combined control over both sound attributes and sounding periods, including counterfactual sound properties, off-screen sounds, and mixtures of counterfactual and normal sound events.
\end{enumerate}

In total, we collected \textbf{100} videos, and for each video we annotated roughly \textbf{1--4} control scenarios per category, yielding \textbf{395} control samples. For evaluation, we use a grounding model to localize sound events. Based on the target sounding periods we annotated, we determine each event’s sounding period and infer the silent intervals. We then measure the overlap between each predicted sound interval (including silence) and the target interval. Segments with an overlap greater than \textbf{0.5} are considered correct predictions, which are then used to compute \textbf{Precision}, \textbf{Recall}, and \textbf{F1-score}.

\noindent \textbf{VGGSound-Director.}
VGGSound-Director focuses on evaluating generation quality. Following the annotation procedure described above, we labeled the STS data on the VGGSound test set and obtained \textbf{2.2K} test samples. On this benchmark, we follow MMAudio and compute audio-quality-related metrics, including distributional quality, perceptual quality, semantics, and audio-visual alignment.

\section{More Results}
\label{sec:results}

\begin{figure*}[ht]
    \centering
    \includegraphics[width=1.0\linewidth]{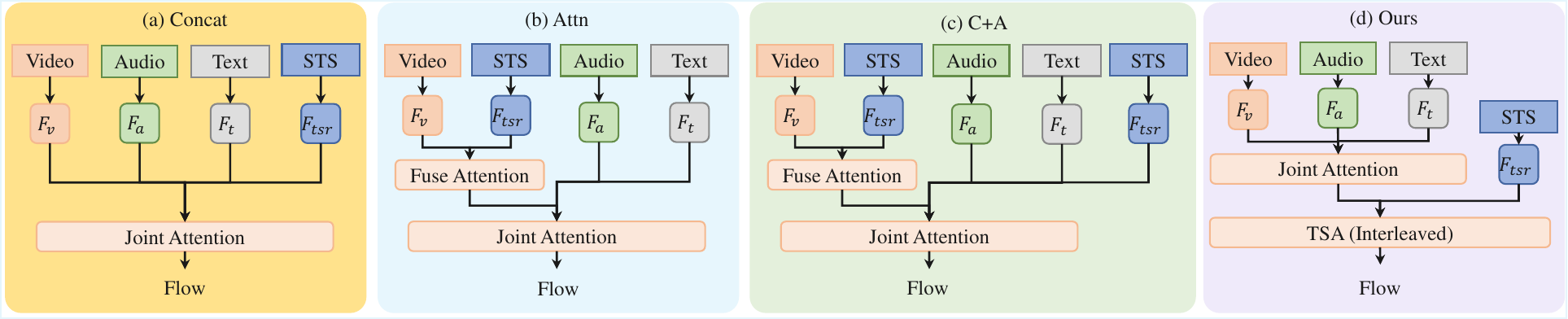}
    \caption{
        \textbf{Different Architecture.} We present the architectural diagrams of different design variants of SG-TFM. (a) concatenates the STS token with features from other modalities; (b) first fuses the STS token with video features using a Fuse-Attention layer; (c) combines the previous two approaches; and (d) shows the design we currently adopt.}
    \label{fig:ablation_arch}
\end{figure*}

\noindent \textbf{DirectorBench.}
As shown in Fig.~\ref{fig:ablation_result_directorbench}, we present a subset of the control results in DirectorBench. Consistent with the main text, we display the mel spectrograms of audio generated by different methods, annotate the target sound categories, and overlay the Grounding model’s confidence for the temporal localization of each sound. The target control time intervals are highlighted above the confidence curves. We observe the following phenomena: 1) Our method improves the accuracy of temporal control. In the leftmost example, we control the time interval for a train horn; other methods either produce a prolonged horn or trigger it at incorrect times. 2) Our method can effectively synthesize off-screen sounds. In the middle example, we aim to create a story of an athlete shouting to motivate themselves before hitting a ball; our method successfully synthesizes the off-screen audio while maintaining a reasonable hitting timing. In the rightmost example, we synthesize background music.

\begin{figure*}[ht]
    \centering
    \includegraphics[width=1.0\linewidth]{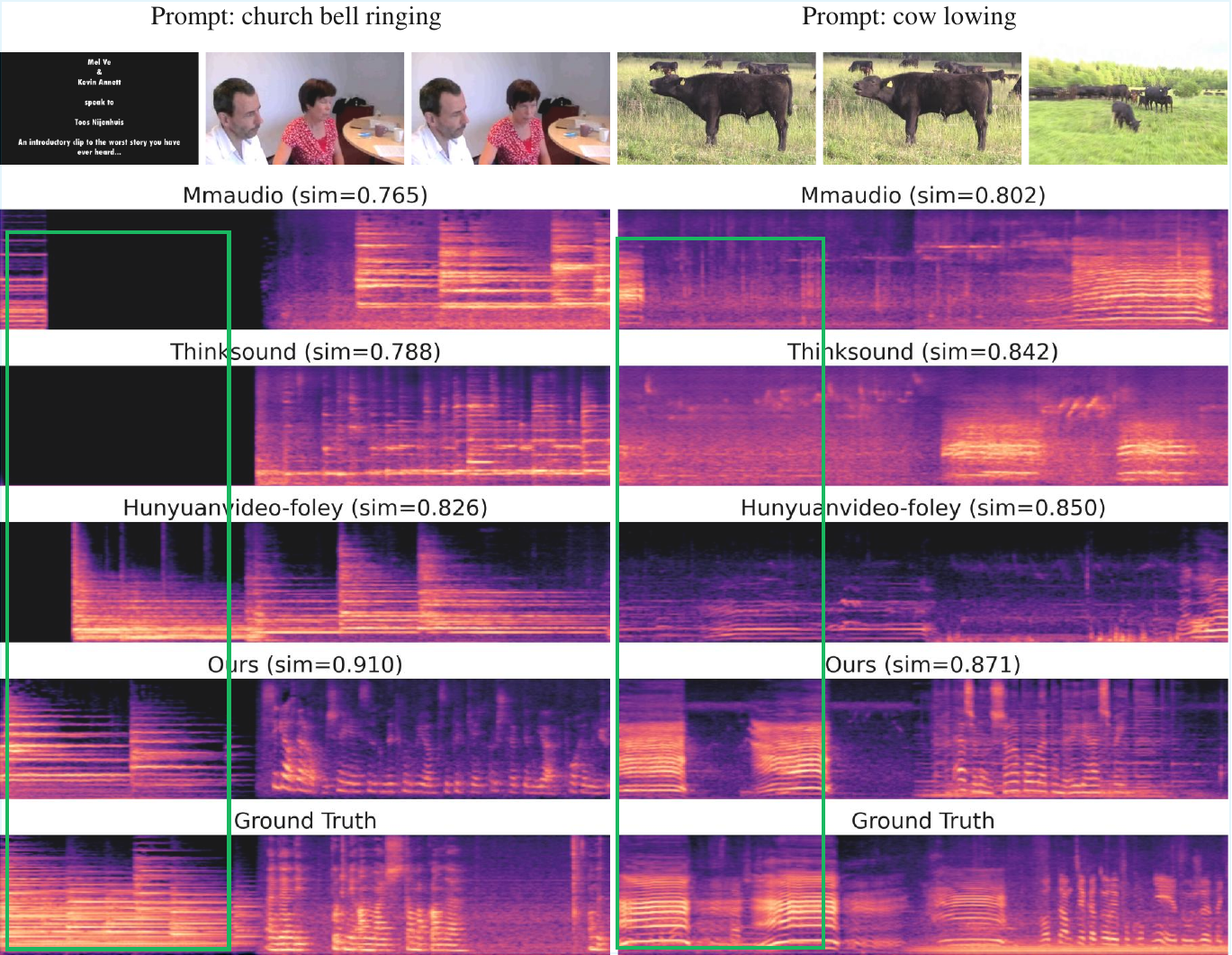}
    \caption{
        \textbf{Visual Results in VGGSound-Director.} We present several results from VGGSound-Director, comparing the mel-spectrograms generated by our method with those from other approaches and with the ground-truth audio. We also compute the L1 similarity between each generated mel-spectrogram and the ground truth.}
    \label{fig:ablation_result_vgg_1}
\end{figure*}

\begin{figure*}[ht]
    \centering
    \includegraphics[width=1.0\linewidth]{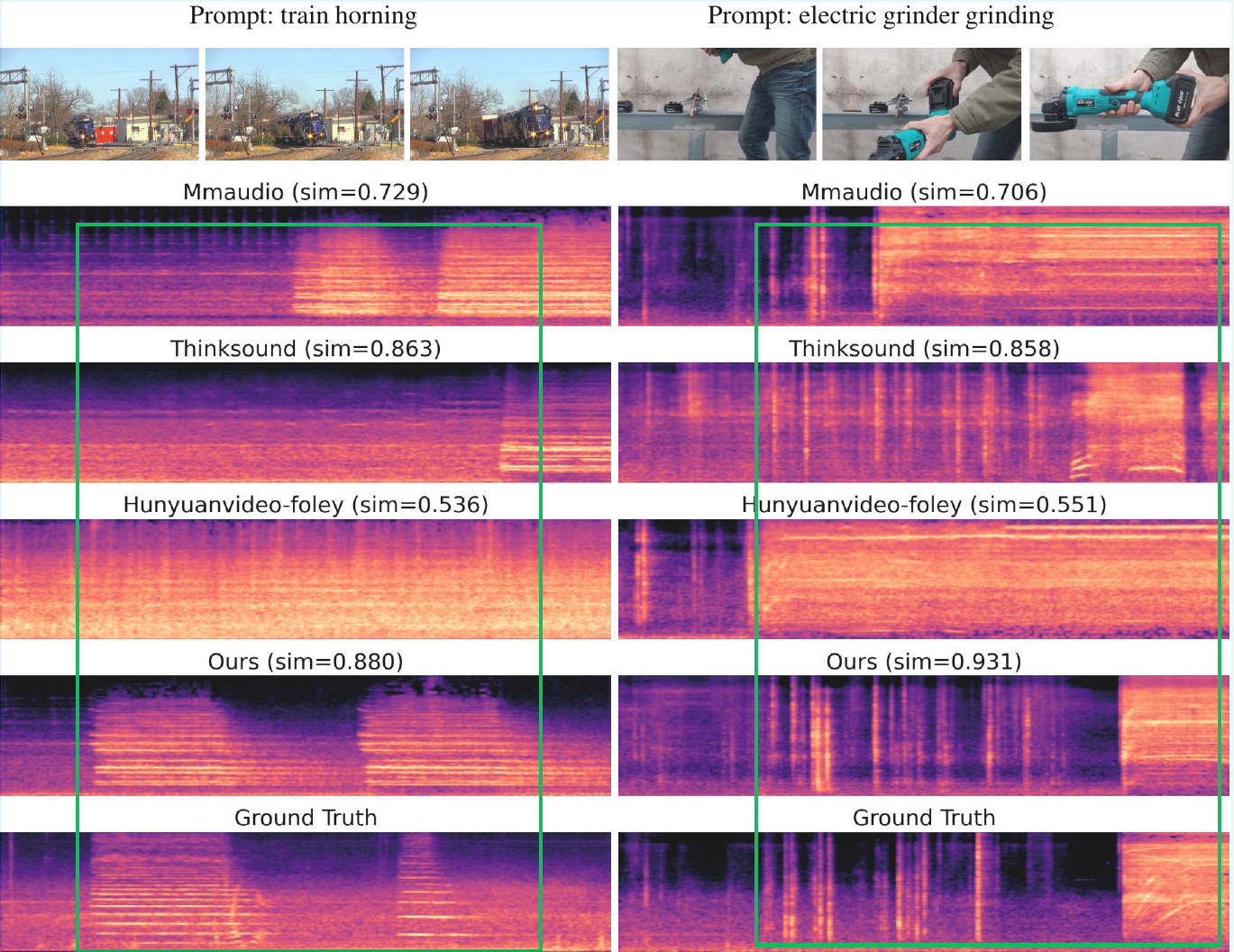}
    \caption{
        \textbf{Visual Results in VGGSound-Director.} We present several results from VGGSound-Director, comparing the mel-spectrograms generated by our method with those from other approaches and with the ground-truth audio. We also compute the L1 similarity between each generated mel-spectrogram and the ground truth.}
    \label{fig:ablation_result_vgg2}
\end{figure*}

\begin{figure*}[ht]
    \centering
    \includegraphics[width=1.0\linewidth]{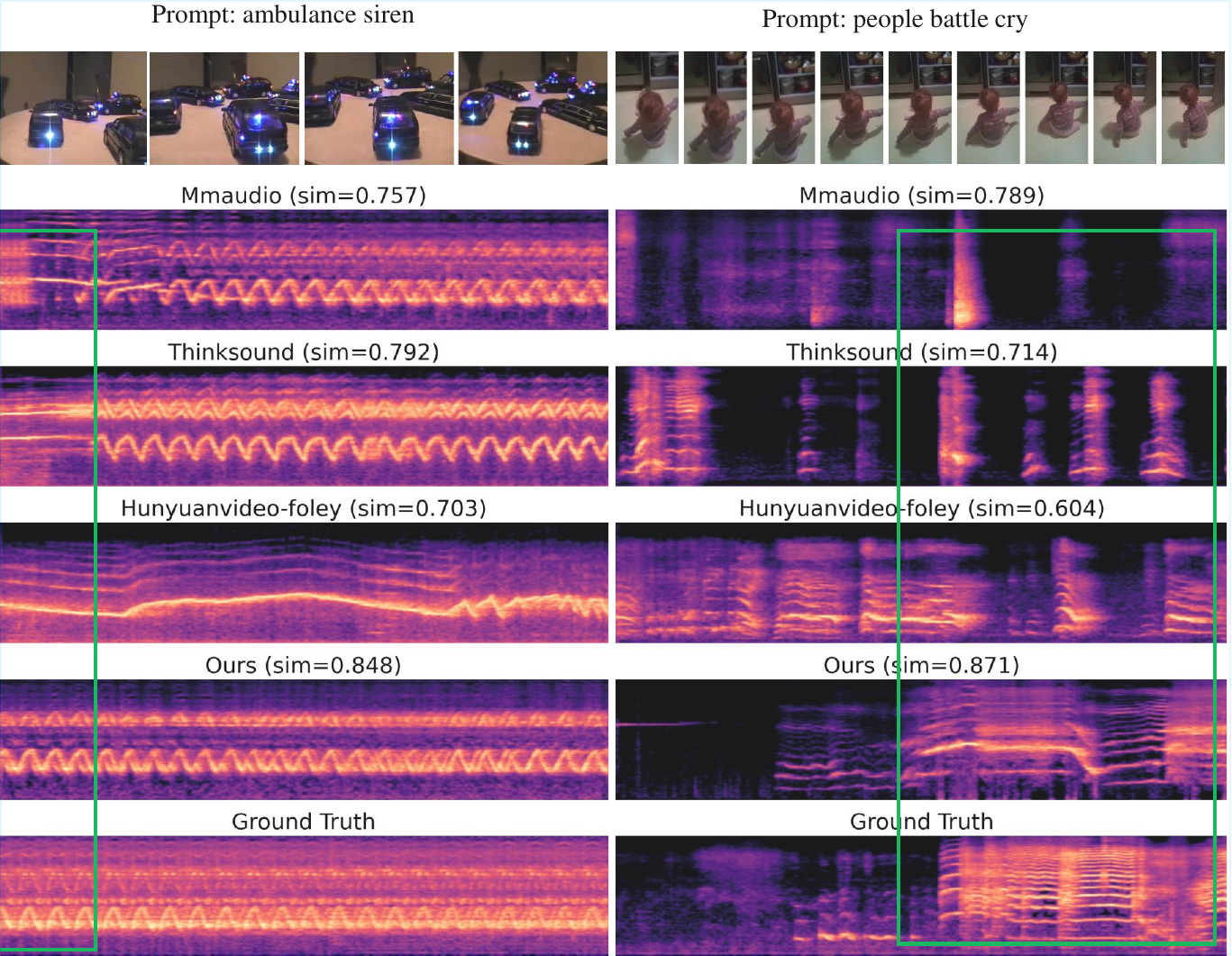}
    \caption{
        \textbf{Visual Results in VGGSound-Director.} We present several results from VGGSound-Director, comparing the mel-spectrograms generated by our method with those from other approaches and with the ground-truth audio. We also compute the L1 similarity between each generated mel-spectrogram and the ground truth.}
    \label{fig:ablation_result_vgg_3}
\end{figure*}

\noindent \textbf{VGGSound-Director.}
We show the mel spectrograms of audio generated by different baselines alongside the mel spectrograms of the corresponding ground-truth audio. We also compute and annotate the L1 similarity (sim) between each method’s mel and the ground-truth mel. The visualization results are presented in Fig.~\ref{fig:ablation_result_vgg_1}, \ref{fig:ablation_result_vgg2} and \ref{fig:ablation_result_vgg_3}. For special videos, such as black screens, text overlays, or transitions, where obvious visual cues are absent, our method synthesizes audio that more closely matches the ground truth, demonstrating its controllability. Additionally, for cases like police sirens or sounds from children facing away from the camera, our method generates audio that is noticeably closer to the real results.

\noindent \textbf{Audio results.}
We have also included in the supplementary materials ZIP a portion of our generated audio results, as well as some outputs from SOTA generators. For each case, we provide a text file describing the control requirements, challenges, and corresponding results.

\section{More Experiments}
\label{experiments}

\noindent \textbf{The Choice of Base-Model.} We evaluated MMAudio models of different sizes (MMAudio-Medium-44K and MMAudio-Large-44K). As shown in the Tab.~\ref{tab:model_size}, although the Large model introduces substantially more parameters and requires greater training resources, it does not yield significant performance improvements. Therefore, we choose to use the more lightweight Medium model for training.

\noindent \textbf{The Impact of Training Iterations.}
The Fig.~\ref{fig:ablation_iters} shows the model performance after different numbers of training iterations. For ease of comparison, all metrics are normalized to the range of 0.3–0.8, with higher values representing better performance—even for metrics such as FD and KL. It can be seen that the red sample (representing 1.2M training iterations) achieves relatively strong performance across almost all metrics. Therefore, we ultimately selected 1.2M iterations, which takes approximately 72 hours.

\renewcommand{\arraystretch}{1.25}

\begin{table}[t]
\small
\centering
\begin{NiceTabular}{@{\hspace{1pt}}>{\centering\arraybackslash}p{1.1cm}
                    @{\hspace{1pt}}c  c c ccc@{\hspace{1pt}}}
\toprule
\textbf{Drop}  & \textbf{FD$_{\text{VGG}}$} $\downarrow$ & \textbf{KL$_{\text{PANN}}$}$\downarrow$ & \textbf{ISC$_{\text{PANN}}$}$\uparrow$ & \textbf{IB}$\uparrow$ & \textbf{DeSync}$\downarrow$ \\
\midrule
0.0  & 1.51 & 1.31 & 14.63 & 0.33 & 0.467 \\
0.1  & 1.17 & 1.42 & 14.84 & 0.33 & 0.432 \\
0.3  & 1.61 & 1.37 & 14.63 & 0.32 & 0.456 \\
0.5  & 1.67 & 1.42 & 14.81 & 0.32 & 0.461 \\
\toprule
\end{NiceTabular}
\caption{Comparison of MMAudio models of different sizes on VGGSound-Director. All models are 44kHz.}
\label{tab:ablation_drop}
\end{table}

\noindent \textbf{The Impact of STS drop ratio.} We evaluated the effect of different STS drop ratios on quality. The results are shown in the Tab.~\ref{tab:ablation_drop}. It can be seen that dropping STS tokens with a probability of 0.1 and replacing them with empty text tokens achieves strong performance across most metrics. Therefore, we adopted this as the default configuration for CFG

\noindent \textbf{The Design of Architecture.} As shown in the Fig.~\ref{fig:ablation_arch}, we tested different architectures for incorporating the STS token. In (a), we concatenate the STS token with features from other modalities and perform the attention operation; in (b), we first fuse STS with video features through an additional attention mechanism to inject temporal information into the video features, and then apply MMAudio’s Joint Attention; in (c), we combine the previous two approaches: the STS token is first fused with video features via attention, and then concatenated with other modality features and the STS features for the attention operation; (d) illustrates the approach we ultimately adopt, where an extra attention layer fuses STS with the already fused audio features. The experimental results are shown in the Tab.~\ref{tab:ablation_arch}. We observe that the concatenation approach achieves lower KL scores but relatively moderate IS and DeSync; the attention-based method achieves the highest ISC but performs poorly on DeSync, likely because the global information in STS partially disrupts the instantaneous temporal properties of video features. The combined approach performs slightly worse than simple concatenation. Finally, our method not only achieves better performance on most metrics but also \textit{allows the SG-TFM module to be optionally discarded to switch between V2A generation and temporal control}, resulting in a more flexible framework with higher overall metrics.


\begin{figure}[ht]
    \centering
    \includegraphics[width=1.0\linewidth]{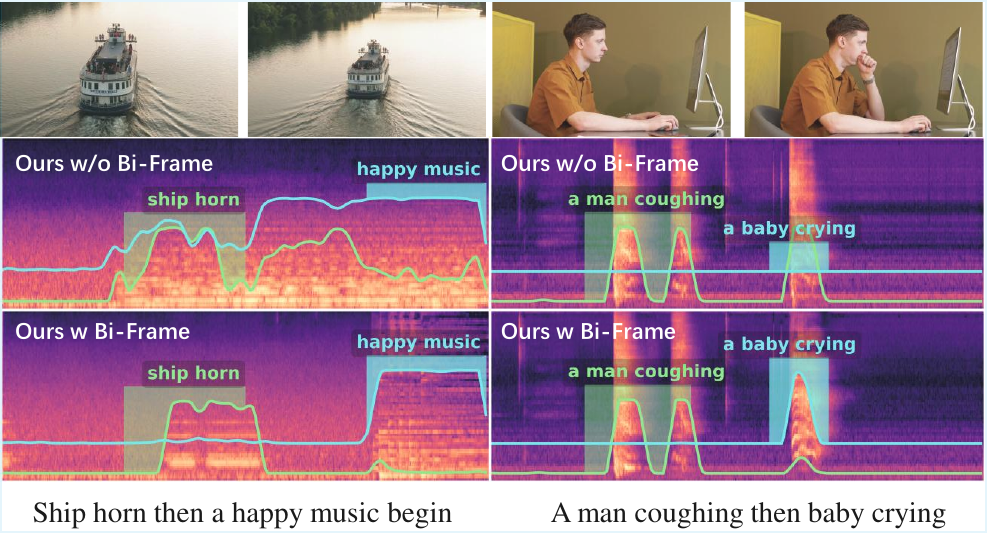}
    \caption{
        \textbf{Visual Results in DirectorBench.} We present several results from DirectorBench.}
    \label{fig:ablation_result_bi}
\end{figure}

\paragraph{The Impact of Bi-Frame.} We also visualized the results of the Bi-Frame framework. It can be observed that without the Bi-Frame strategy, when synthesizing certain off-screen or counterfactual sounds, the corresponding audio in our method is affected by visual information, leading to weaker control or attribute leakage. By employing the Bi-Frame framework, this issue is flexibly addressed, improving both controllability and flexibility.

\section{Training Details}

We train our model for \textbf{1200K} iterations with a batch size of \textbf{16}. 
We train our model on 8×A800 GPUs (40 GB each) using the MMAudio-medium architecture, with a total training time of approximately 72 hours. During training, we randomly drop TSR features with a probability of 0.1
We set an initial learning rate of \textbf{$2.0\times10^{-5}$}. We use the AdamW optimizer and a cosine learning rate decay schedule.
We apply a weight decay of \textbf{$1.0\times10^{-6}$} and clip gradients at a maximum norm of \textbf{1.0}. 
For training efficiency, we use bf16 mixed precision training, and all the audio latents and visual embeddings are
precomputed offline and loaded during training.

\renewcommand{\arraystretch}{1.25}

\begin{table}[t]
\small
\centering
\begin{NiceTabular}{@{\hspace{1pt}}>{\centering\arraybackslash}p{1.1cm}
                    @{\hspace{1pt}}c  c c ccc@{\hspace{1pt}}}
\toprule
\textbf{Arch}  & \textbf{FD$_{\text{VGG}}$} $\downarrow$ & \textbf{KL$_{\text{PANN}}$}$\downarrow$ & \textbf{ISC$_{\text{PANN}}$}$\uparrow$ & \textbf{IB}$\uparrow$ & \textbf{DeSync}$\downarrow$ \\
\midrule
Concat & 1.29 & 1.26 & 14.62 & 0.33 & 0.476 \\
Attn  & 1.74 & 1.54 & 15.49 & 0.32 & 0.535 \\
C+A  & 1.43 & 1.30 & 14.59 & 0.32 & 0.476 \\
Ours  & 1.17 & 1.42 & 14.84 & 0.33 & 0.432 \\
\toprule
\end{NiceTabular}
\caption{Comparison of MMAudio models of different sizes on VGGSound-Director. All models are 44kHz.}
\label{tab:ablation_arch}
\end{table}

\section{Limitations and Future work}
\noindent \textbf{Limitation.} 
Although our approach enhances the controllability of the generation process, it still has several limitations. \textbf{1)} Our framework is built upon the MMAudio architecture, and thus its performance is inherently constrained by the limitations of MMAudio itself. \textbf{2)} Since the training audio in current V2A datasets is relatively simple, our model struggles to produce satisfactory results when synthesizing highly complex audio, due to both data limitations and model capacity. 
\noindent \textbf{Future Work.} 
In traditional V2A tasks, the audio data and its annotations are relatively simple. This means that conventional V2A methods focus on a small set of audio event combinations, with only a single audio sound per time segment. We argue that though our method also only focus on the single event per segment, \textbf{the decoupling concept underlying Bi-Frame also holds potential for addressing multi-audio multi-track event combinations}. For example, one could replace Bi-Frame’s off-screen/on-screen decoupling with a decoupling of individual audio events. For each audio event, a separate track is assigned and the audio is synthesized under STS control. During fusion, we can adjust the fusion weight of each track within the segment to achieve multi-event, multi-track synthesis.

\end{document}